\documentclass[aps,prd,twocolumn,superscriptaddress,nofootinbib,floatfix]{revtex4-2}

\usepackage[T1]{fontenc}
\usepackage[utf8]{inputenc}
\usepackage{lmodern}
\usepackage[reqno]{amsmath}    
\usepackage{bm,amssymb,mathtools}
\usepackage{graphicx}
\usepackage{pgfplots}
\pgfplotsset{compat=1.18}
\usepackage{hyperref}
\hypersetup{colorlinks=true,linkcolor=blue,citecolor=blue,urlcolor=blue}
\usepackage{placeins}
\usepackage{booktabs}

\newcommand{\e}{\epsilon}      
\newcommand{\Bpar}{B}
\newcommand{\ordone}{\mathcal{O}(1)}
\newcommand{\arrtight}{\setlength{\arraycolsep}{4pt}\renewcommand{\arraystretch}{1.05}}

\begin{document}

\title{One-Flavon Flavor: A Single \texorpdfstring{\emph{Hierarchical}}{Hierarchical}
Parameter \texorpdfstring{$B$}{B} Organizes Quarks \texorpdfstring{\&}{and}
Leptons at \texorpdfstring{$M_Z$}{M\_Z}}

\author{Vernon Barger}
\affiliation{Department of Physics, University of Wisconsin--Madison, Madison, WI 53706, USA}

\date{\today}

\begin{abstract}
Flavor hierarchies emerge from a single hierarchical parameter $B$ in a one-flavon Froggatt--Nielsen scheme. Fixing $B=5.357$ from charged-lepton ratios ($m_e:m_\mu:m_\tau\!\propto\!\e^5:\e^2:1$, $\e=1/B$), we reproduce quark masses and CKM targets at $M_Z$ with $\ordone$ coefficients. The same $\e$ gives viable lepton textures and benchmarks for $U_{\rm PMNS}$, $\sum m_\nu$, and $m_{\beta\beta}$. Compact correlations ($|V_{us}|,|V_{cb}|,|V_{ub}|,\theta_{13}^{\rm PMNS}$) follow from powers of $\e$.
\end{abstract}

\maketitle

\section{Introduction}

The observed hierarchies of quark and lepton masses and mixings span many orders of magnitude, yet are accurately described by the Standard Model (SM) Yukawa sector with apparently arbitrary $\ordone$ complex parameters.  Froggatt--Nielsen (FN) models \cite{Froggatt:10.1016/0550-3213(79)90262-3} offer a simple organizing principle: a spontaneously broken $U(1)_{\rm FN}$ symmetry and one or more flavon fields generate hierarchical Yukawas via powers of a small parameter $\e = \langle \phi \rangle / \Lambda$.

In this work we show that a \emph{single} hierarchical parameter $B$---fixed by charged-lepton ratios---organizes the entire set of quark and lepton hierarchies at $M_Z$ in a one-flavon FN scheme. Concretely, we take
\begin{equation}
\label{eq:B_def}
\Bpar \equiv \frac{1}{\e},\qquad
m_e:m_\mu:m_\tau \propto \e^{5}:\e^{2}:1,
\end{equation}
and fit $B$ from charged-lepton data.\footnote{A numerical derivation of $\e$ from $m_e/m_\tau$ is given in Sec.~\ref{subsec:eps_determination}.}
With $B=5.357$ ($\e=0.187$) we obtain simple integer FN exponents for the Yukawa textures and reproduce quark masses and the leading CKM/PMNS structures with near-unity complex coefficients.

We work with a minimal one-flavon $U(1)_{\rm FN}$ framework and provide an explicit benchmark set of complex $\ordone$ matrices that reproduces:
\begin{itemize}
  \item quark masses at $M_Z$ (Table~\ref{tab:quarkmasses}),
  \item CKM magnitudes and phase (Table~\ref{tab:ckm}, Eqs.~\eqref{eq:Vckm_phaseful}--\eqref{eq:argVckm}),
  \item PMNS mixing angles and a benchmark leptonic Dirac phase (Eqs.~\eqref{eq:Unu_best}, \eqref{eq:argUpmns}),
\end{itemize}
for a fixed FN charge choice taken from the scan of Ref.~\cite{Cornella:2023qpr}.  We also show explicitly how the extracted $\ordone$ coefficients behave in a Type--II 2HDM interpretation with $\tan\beta=40$ and in the single--Higgs SM.

\section{FN framework and exponents}
\label{sec:FNframework}

\subsection{FN Lagrangian}
\label{subsec:FN_Lagrangian}

We work with a Froggatt--Nielsen $U(1)_{\rm FN}$ symmetry with one flavon $\phi$ of charge $q(\phi)=-1$, and a single Higgs doublet $H$ with $q(H)=0$.  We define
\begin{equation}
\e \equiv \frac{\langle\phi\rangle}{\Lambda} = \frac{1}{\Bpar},
\qquad
\Bpar = 5.357\;\Rightarrow\;\e \simeq 0.187,
\end{equation}
so that Yukawa entries scale as integral powers of $\e$.

The FN expansion of the Yukawa and Weinberg operators is
\begin{equation}
\begin{aligned}
\mathcal{L}_{\rm FN}\supset {} &
\sum_{ij}\lambda^u_{ij}\!\left(\frac{\phi}{\Lambda}\right)^{n^u_{ij}} \bar Q_i H u^c_j
+\sum_{ij}\lambda^d_{ij}\!\left(\frac{\phi}{\Lambda}\right)^{n^d_{ij}} \bar Q_i \tilde H d^c_j
\\
&+\sum_{ij}\lambda^e_{ij}\!\left(\frac{\phi}{\Lambda}\right)^{n^e_{ij}} \bar L_i \tilde H e^c_j
\\
&+\frac{1}{\Lambda_L}\sum_{ij}\kappa_{ij}\!\left(\frac{\phi}{\Lambda}\right)^{n^\nu_{ij}}
(\bar L_i \tilde H)(\bar L_j \tilde H)
\;+\;{\rm h.c.}
\end{aligned}
\label{eq:FNlag}
\end{equation}
with $\tilde H=i\sigma_2 H^*$ and integer exponents $n^{f}_{ij}$ determined by the $U(1)_{\rm FN}$ charges.  After electroweak symmetry breaking (EWSB) with $\langle H\rangle=v/\sqrt{2}$, the Higgs vev sets the overall mass scale, while the \emph{flavon} controls hierarchies and mixings through powers of $\e$.

In 2HDM/MSSM (Type-II) realizations, one replaces $H\to(H_u,H_d)$ and $v\to(v_u,v_d)$ with $\tan\beta\equiv v_u/v_d$.  In that case “factorize overall scales’’ means the fermion mass matrices pick up overall factors of $v_u$ or $v_d$ (depending on the sector), while the FN \emph{exponents}, and hence the hierarchical structure, remain unchanged.

\subsection{FN charges and exponents}
\label{subsec:charges_exponents}

The exponents $n^f_{ij}$ are fixed by the $U(1)_{\rm FN}$ charges of the fields, which we take from Ref.~\cite{Cornella:2023qpr} as a simple viable assignment:
\begin{table}[t]
\caption{FN charges used in this work (quarks and two lepton options).}
\begin{ruledtabular}
\begin{tabular}{lcc}
Sector & Field & Charges \\
\hline
Quarks & $Q_i$ & $(3,2,0)$ \\
       & $u^c_j$ & $(4,1,0)$ \\
       & $d^c_j$ & $(1,0,0)$ \\
\hline
Leptons (A) & $L_i$ & $(1,0,0)$ \\
            & $e^c_j$ & $(4,2,0)$ \\
Leptons (B) & $L_i$ & $(0,1,0)$ \\
            & $e^c_j$ & $(5,1,0)$ \\
\end{tabular}
\end{ruledtabular}
\label{tab:FNcharges}
\end{table}

The FN exponents are obtained by demanding $U(1)_{\rm FN}$ invariance of each operator, i.e.
\begin{itemize}
    \item Up-type quarks: $\bar Q_i H u^c_j$ gives $n^u_{ij} = q(Q_i) + q(u^c_j)$,
    \item Down-type quarks: $\bar Q_i \tilde H d^c_j$ gives $n^d_{ij} = q(Q_i) + q(d^c_j)$,
    \item Charged leptons: $\bar L_i \tilde H e^c_j$ gives $n^e_{ij} = q(L_i) + q(e^c_j)$,
    \item Majorana neutrinos (Weinberg operator): $(\bar L_i \tilde H)(\bar L_j \tilde H)$ gives $n^\nu_{ij} = q(L_i) + q(L_j)$.
\end{itemize}
For example, the $(1,1)$ entry of $Y_u$ has exponent
\begin{equation}
n^u_{11} = q(Q_1) + q(u^c_1) = 3 + 4 = 7,
\end{equation}
in agreement with Eq.~\eqref{eq:YuYd} below.

For the quark and charged-lepton diagonal entries we will refer to the exponents
\begin{equation}
p_{u,c,t} = (7,3,0),\quad
p_{d,s,b} = (4,2,0),\quad
p_{e,\mu,\tau} = (5,2,0),
\end{equation}
while for the mixing observables we take
\begin{equation}
p_{us}=1,\quad p_{cb}=2,\quad p_{ub}=3,\quad p_{13}=1,
\end{equation}
so that $|V_{us}|\sim \e$, $|V_{cb}|\sim \e^2$, $|V_{ub}|\sim \e^3$, and $\theta_{13}^{\rm PMNS}\sim \e$.

\subsection{Determination of the hierarchical parameter \texorpdfstring{$\e$}{epsilon}}
\label{subsec:eps_determination}

The hierarchical parameter $\e \equiv 1/B$ is fixed by charged-lepton ratios, which we assume to scale as
\begin{equation}
m_e:m_\mu:m_\tau \propto \e^5:\e^2:1.
\end{equation}
The most robust ratio is $m_e/m_\tau$, spanning the full charged-lepton hierarchy.  Using RGE-run values at $M_Z$ (approximately $m_e(M_Z)\approx 0.48$ MeV and $m_\tau(M_Z)\approx 1728$ MeV), we obtain
\begin{equation}
\frac{m_e(M_Z)}{m_\tau(M_Z)}
\;\approx\;
\frac{0.48~\mathrm{MeV}}{1728~\mathrm{MeV}}
\;\approx\; 2.78\times 10^{-4}.
\end{equation}
Setting this equal to the model scaling $\e^5$ gives
\begin{equation}
\e^5 \simeq 2.78\times 10^{-4}
\quad\Rightarrow\quad
\e \simeq (2.78\times 10^{-4})^{1/5} \approx 0.194,
\end{equation}
very close to the benchmark value $\e=0.187$ used throughout (corresponding to $B=5.357$).  The quoted value results from a more precise global fit across all observables, but is dominantly fixed by the charged-lepton hierarchy.

\section{Textures and benchmark coefficient matrices}
\label{sec:textures}

\subsection{FN textures}
\label{subsec:FN_textures}

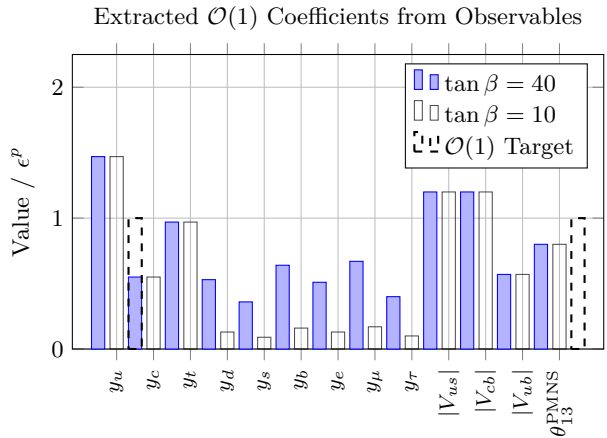
\begin{figure}[!tbp]
\centering
\begin{tikzpicture}
\begin{axis}[
    ybar,
    width=\columnwidth,
    height=5.5cm,
    bar width=5pt,
    title={Extracted $\mathcal{O}(1)$ Coefficients from Observables},
    ylabel={Value / $\e^p$},
    ymin=0, ymax=2.25,
    symbolic x coords={
        yu, yc, yt,
        yd, ys, yb,
        ye, ymu, ytau,
        Vus, Vcb, Vub,
        theta13
    },
    xtick=data,
    xticklabels={
        $y_u$, $y_c$, $y_t$,
        $y_d$, $y_s$, $y_b$,
        $y_e$, $y_\mu$, $y_\tau$,
        $|V_{us}|$, $|V_{cb}|$, $|V_{ub}|$,
        $\theta_{13}^{\rm PMNS}$
    },
    x tick label style={rotate=90, anchor=east, font=\footnotesize},
    grid=major,
    legend pos=north east,
    legend cell align={left},
    legend columns=1
]
\addplot coordinates {
    (yu, 1.47) (yc, 0.55) (yt, 0.97)
    (yd, 0.53) (ys, 0.36) (yb, 0.64)
    (ye, 0.51) (ymu, 0.67) (ytau, 0.40)
    (Vus, 1.20) (Vcb, 1.20) (Vub, 0.57)
    (theta13, 0.80)
};
\addlegendentry{$\tan\beta=40$}

\addplot[opacity=0.7] coordinates {
    (yu, 1.47) (yc, 0.55) (yt, 0.97)
    (yd, 0.13) (ys, 0.09) (yb, 0.16)
    (ye, 0.13) (ymu, 0.17) (ytau, 0.10)
    (Vus, 1.20) (Vcb, 1.20) (Vub, 0.57)
    (theta13, 0.80)
};
\addlegendentry{$\tan\beta=10$}

\addplot [black, dashed, thick] coordinates {(yu,1) (theta13,1)};
\addlegendentry{$\mathcal{O}(1)$ Target}

\end{axis}
\end{tikzpicture}
\caption{
Underlying $\mathcal{O}(1)$ coefficients, extracted by dividing observables by their predicted FN scaling power $p$.  Masses are converted to Yukawas $y_f = m_f/v_f$ assuming a Type-II 2HDM, with $v_{\rm SM}=174.1$ GeV.  Up-sector coefficients $\{y_u,y_c,y_t\}$ (using $v_u \approx v_{\rm SM}$) and mixing coefficients $\{|V_{ij}|,\theta_{13}\}$ are independent of $\tan\beta$.  Down-sector and lepton coefficients $\{y_d,y_s,y_b,y_e,y_\mu,y_\tau\}$ scale as $\approx \tan\beta$ (since $y_f = m_f/v_d$ and $v_d \simeq v_{\rm SM}/\tan\beta$).  The plot shows that $\tan\beta=40$ keeps these coefficients $\ordone$, while $\tan\beta=10$ suppresses them to $\sim 0.1$.  Powers $p$ are
$\{y_u,y_c,y_t\}:\{7,3,0\}$,
$\{y_d,y_s,y_b\}:\{4,2,0\}$,
$\{y_e,y_\mu,y_\tau\}:\{5,2,0\}$,
$\{|V_{us}|,|V_{cb}|,|V_{ub}|,\theta_{13}^{\rm PMNS}\}:\{1,2,3,1\}$.
For the single--Higgs SM baseline, one instead takes $v_f=v_{\rm SM}$ for all fermions; the down-sector and lepton coefficients are then uniformly reduced by a factor $\sim 1/40$, leaving the hierarchy of exponents unchanged (see Appendix~\ref{app:coeffs}).}
\label{fig:o1plot}
\end{figure}

The quark Yukawa textures implied by the exponents discussed above can be written as
\begin{equation}\label{eq:YuYd}
\begin{aligned}
Y_u &\sim C_u \circ
\begin{pmatrix}
\e^{7} & \e^{4} & \e^{3}\\
\e^{6} & \e^{3} & \e^{2}\\
\e^{4} & \e^{1} & 1
\end{pmatrix},
\\[4pt]
Y_d &\sim C_d \circ
\begin{pmatrix}
\e^{4} & \e^{3} & \e^{3}\\
\e^{3} & \e^{2} & \e^{2}\\
\e^{1} & 1      & 1
\end{pmatrix},
\end{aligned}
\end{equation}
where “$\circ$’’ denotes the Hadamard (elementwise) product.  The complex matrices $C_u=(C^{u}_{ij})$ and $C_d=(C^{d}_{ij})$ collect the residual $\ordone$ coefficients, including CP phases; their magnitudes are expected to lie in an $\ordone$ band (see Fig.~\ref{fig:o1plot} and Appendix~\ref{app:coeffs}).

In the benchmark fit we work in a basis where the up Yukawa is diagonal, so all CKM mixing arises from $Y_d$.  This choice is purely conventional; only the relative left-handed rotations of $Y_u$ and $Y_d$ are physical.

\subsection{Extracted \texorpdfstring{$\ordone$}{O(1)} coefficients}
\label{subsec:coeff_plot}

To visualize the FN structure, we extract effective \(\ordone\) coefficients from observables by dividing out the predicted power of $\e$:
\begin{equation}
c_f \equiv \frac{y_f}{\e^{p_f}},\qquad
c_{V_{ij}} \equiv \frac{|V_{ij}|}{\e^{p_{ij}}},\qquad
c_{\theta_{13}} \equiv \frac{\theta_{13}^{\rm PMNS}}{\e^{p_{13}}}.
\end{equation}
Here $y_f$ are the Yukawa couplings obtained from quark and charged-lepton masses at $M_Z$, and we use the exponents $p_f$ and $p_{ij}$ defined above.  For illustration we interpret the down-type Yukawas as Type-II 2HDM Yukawas with $\tan\beta=40$, so that $y_{d,\ell}=m_{d,\ell}/v_d$ with $v_d\simeq v_{\rm SM}/40$; the up-sector Yukawas use $v_u\simeq v_{\rm SM}$.  (The single--Higgs SM reinterpretation is discussed in Appendix~\ref{app:coeffs}.)

\subsection{Benchmark $C$ matrices}
\label{subsec:benchmarkCs}

To close the reproducibility gap we provide an explicit benchmark set of \(\ordone\) complex matrices $C_u$ and $C_d$ that, together with Eq.~\eqref{eq:YuYd} at $\e=0.187$, exactly reproduce the target quark masses in Table~\ref{tab:quarkmasses} and the CKM magnitudes in Table~\ref{tab:ckm} (up to trivial rephasings and rounding), when interpreted in a Type--II 2HDM with $\tan\beta=40$.

We work in a basis where the up Yukawa matrix is diagonal, so that all CKM mixing arises from the down sector.  In this basis one may take
\begin{equation}
Y_u = \mathrm{diag}(y_u,y_c,y_t)
= \mathrm{diag}(m_u,m_c,m_t)\,\frac{\sqrt{2}}{v},
\end{equation}
with $v=174.1$ GeV, and enforce
\begin{equation}
Y_d = V_{\rm CKM}\,\mathrm{diag}(y_d,y_s,y_b),
\end{equation}
where $V_{\rm CKM}$ is taken in the PDG parameterization with
\[
|V_{us}|=0.225,\quad |V_{cb}|=0.0418,\quad |V_{ub}|=0.00372,\quad
\delta_{\rm CKM}=65.4^\circ.
\]
The down Yukawas are interpreted as Type-II couplings,
$y_{d,s,b} = m_{d,s,b}/v_d$ with $v_d\simeq v/\tan\beta$ and $\tan\beta=40$.

The benchmark $C$ matrices that implement this construction are:
\arrtight
\begin{subequations}\label{eq:BenchmarkCs}
\begin{equation}\label{eq:Cu_bench}
C_u \approx
\begin{pmatrix}
0.797 & 0     & 0     \\
0     & 0.552 & 0     \\
0     & 0     & 0.971
\end{pmatrix}
\end{equation}
and $C_d=$
\begin{equation}\label{eq:Cd_bench}
\begin{pmatrix}
0.516 & 0.440 & 0.365\,e^{-\,i\,1.14}\\[3pt]
0.0223\,e^{+\,i\,\pi} & 0.356 & 0.768\\[3pt]
2.97\times 10^{-5}\,e^{-\,i\,0.395} &
5.26\times 10^{-4}\,e^{+\,i\,\pi} &
0.642
\end{pmatrix}
\end{equation}
For completeness, we retain the illustrative neutrino-sector coefficient matrix
\begin{equation}\label{eq:K_bench}
K \approx
\begin{pmatrix}
1.02\,e^{+i\,0.40} & 0.97\,e^{-i\,0.50} & 1.05\,e^{+i\,0.90}\\
0.97\,e^{-i\,0.50} & 1.03\,e^{+i\,0.20} & 0.99\,e^{-i\,0.65}\\
1.05\,e^{+i\,0.90} & 0.99\,e^{-i\,0.65} & 0.96\,e^{+i\,0.10}
\end{pmatrix},
\end{equation}
which is used in the Majorana mass matrix $M_\nu$ below.
\end{subequations}
\noindent All entries are quoted to three significant figures in magnitude (and in phase, where applicable).  In particular, the diagonal elements of $C_u$ are chosen such that
\[
y_u=C_{u,11}\e^7,\quad
y_c=C_{u,22}\e^3,\quad
y_t=C_{u,33},
\]
reproducing the up-type Yukawas from Table~\ref{tab:quarkmasses}.  The down-sector matrix $C_d$ is obtained by dividing the constructed $Y_d$ by the FN powers in Eq.~\eqref{eq:YuYd}; its entries are all comfortably within the broad $\ordone$ band.

\subsection{Diagonalization and CKM power counting}
\label{subsec:diag_scaling}

Let $U^{f}_L, U^{f}_R$ be the unitary matrices that diagonalize the Yukawas,
\begin{equation}
U^{f\dagger}_L Y_f U^{f}_R=\mathrm{diag}(y_{f_1},y_{f_2},y_{f_3})\,,\qquad f=u,d.
\end{equation}
The CKM matrix is $V_{\rm CKM}=U_L^{u\dagger}U_L^{d}$.

Ignoring $\ordone$ coefficients, the FN texture in Eq.~\eqref{eq:YuYd} implies the scaling
\begin{subequations}\label{eq:ULscales}
\begin{equation}\label{eq:ULu_scaling}
U_L^{u} \sim
\begin{pmatrix}
1 & \ordone\,\e & \ordone\,\e^3\\
\ordone\,\e & 1 & \ordone\,\e^2\\
\ordone\,\e^3 & \ordone\,\e^2 & 1
\end{pmatrix},
\end{equation}
\begin{equation}\label{eq:ULd_scaling}
U_L^{d} \sim
\begin{pmatrix}
1 & \ordone\,\e & \ordone\,\e^3\\
\ordone\,\e & 1 & \ordone\,\e^2\\
\ordone\,\e^3 & \ordone\,\e^2 & 1
\end{pmatrix}.
\end{equation}
\end{subequations}
Thus at leading order one finds
\begin{equation}
|V_{us}|\sim\e,\quad
|V_{cb}|\sim\e^2,\quad
|V_{ub}|\sim\e^3,
\end{equation}
which is borne out numerically by the benchmark fit.

For the explicit benchmark $C$ matrices in Eq.~\eqref{eq:BenchmarkCs}, we find
\begin{subequations}\label{eq:UL_best}
\begin{equation}\label{eq:ULu_best}
U_L^{u} \approx \mathbb{I},
\end{equation}
\begin{equation}\label{eq:ULd_best}
U_L^{d} \approx
\begin{pmatrix}
 0.974  & 0.225  & 0.00372\\
 -0.225 & 0.974  & 0.0418\\
 0.00670& -0.0414& 0.999
\end{pmatrix},
\end{equation}
\end{subequations}
to three significant figures.  This realizes the standard picture where CKM mixing is dominantly sourced by the down sector.  Right rotations are basis dependent and unobservable in charged currents; in the benchmark we take $U_R^{u}\approx\mathbb{I}$ and $U_R^{d}\approx\mathbb{I}$.

\section{Lepton sector and PMNS}
\label{sec:leptons}

\subsection{Neutrino sector and Takagi factorization}

We assume neutrino masses are generated by the FN-suppressed Weinberg operator in Eq.~\eqref{eq:FNlag}.  The resulting Majorana mass matrix can be written schematically as
\begin{equation}
M_\nu \sim K \circ \left(\e^{n^\nu_{ij}}\right) \frac{v^2}{\Lambda_L},
\end{equation}
with $K$ an $\ordone$ complex symmetric matrix.  For Majorana masses there is no independent $U_R^\nu$; instead a Takagi factorization diagonalizes $M_\nu$:
\begin{equation}
U_\nu^{T} M_\nu\, U_\nu=\mathrm{diag}(m_1,m_2,m_3),\qquad m_i\ge 0.
\end{equation}
The lepton mixing matrix is then
\begin{equation}
U_{\rm PMNS}=U_e^\dagger U_\nu,
\end{equation}
where $U_e$ diagonalizes the charged-lepton Yukawa matrix $Y_e$ (not shown explicitly).

For lepton option A in Table~\ref{tab:FNcharges}, with $L_i$ charges $(1,0,0)$, the neutrino exponents are
\begin{equation}
n^\nu_{ij} = q(L_i)+q(L_j)
=
\begin{pmatrix}
2 & 1 & 1\\
1 & 0 & 0\\
1 & 0 & 0
\end{pmatrix},
\end{equation}
so that
\begin{equation}
\label{eq:Mnu_texture_eps}
M_\nu \sim \frac{v^2}{\Lambda_L}\,
K \circ
\begin{pmatrix}
\e^2 & \e & \e\\
\e   & 1 & 1\\
\e   & 1 & 1
\end{pmatrix},
\end{equation}
with each matrix element scaling as $\e^{n^\nu_{ij}}=B^{-\,n^\nu_{ij}}$.

For the benchmark matrix $K$ in Eq.~\eqref{eq:K_bench} and suitable charged-lepton rotations, we obtain the neutrino mixing moduli
\begin{equation}\label{eq:Unu_best}
|U_\nu| \approx
\begin{pmatrix}
0.825 & 0.545 & 0.149\\
0.309 & 0.588 & 0.747\\
0.472 & 0.595 & 0.649
\end{pmatrix},
\end{equation}
corresponding to
\begin{equation}
(\theta_{12},\theta_{23},\theta_{13})\simeq(33.4^\circ,49.0^\circ,8.60^\circ),
\end{equation}
in good agreement with current global fits.  The Dirac and Majorana phases do not affect these moduli.

\subsection{Extraction of Dirac phases}
\label{subsec:phase_extraction}

For completeness we summarize how the Dirac CP phases are extracted from the benchmark mixing matrices.

\paragraph*{CKM phase from \boldmath$V_{td}$.}
We adopt the standard PDG parameterization of the CKM matrix with three angles
$(\theta_{12},\theta_{23},\theta_{13})$ and one CP phase $\delta$.
At the precision relevant here we may take
\[
s_{12}\equiv \sin\theta_{12} \simeq |V_{us}|,\quad
s_{23}\simeq |V_{cb}|,\quad
s_{13}\simeq |V_{ub}|,
\]
and \(c_{ij}\equiv\sqrt{1-s_{ij}^2}\).  In this parameterization the magnitude of $V_{td}$ is
\begin{equation}
\label{eq:Vtd_exact}
|V_{td}| \;=\; \bigl|\, s_{12}s_{23} \;-\; c_{12}c_{23}s_{13}\,e^{i\delta}\,\bigr|\,.
\end{equation}
Squaring and solving for $\cos\delta$ gives
\begin{equation}
\label{eq:cosdelta}
\cos\delta \;=\;
\frac{(s_{12}s_{23})^2 + (c_{12}c_{23}s_{13})^2 - |V_{td}|^2}
     {2\,s_{12}s_{23}\,c_{12}c_{23}\,s_{13}}\,.
\end{equation}
Using the benchmark values $s_{12}=0.225$, $s_{23}=0.0418$, $s_{13}=0.00372$, and $|V_{td}|=0.00856$ yields
\begin{equation}
\label{eq:Vtd_delta_num}
\begin{aligned}
|V_{td}| &\simeq 0.00856,\\
\delta_{\rm CKM} &\simeq 65.4^\circ.
\end{aligned}
\end{equation}

\paragraph*{PMNS phase from unitarity.}
For the leptons we adopt the PDG parameterization of $U_{\rm PMNS}$ with three angles
$(\theta_{12},\theta_{23},\theta_{13})$ and a Dirac phase $\delta_{\rm PMNS}$.
Defining $s_{ij}\equiv\sin\theta_{ij}$, $c_{ij}\equiv\cos\theta_{ij}$, the elements $U_{\mu 1}$ and $U_{\tau 1}$ take the form
\begin{align}
\label{eq:Umu1_def}
U_{\mu 1} &= -\,s_{12}c_{23}\;-\;c_{12}s_{23}s_{13}\,e^{i\delta_{\rm PMNS}},\\
\label{eq:Utau1_def}
U_{\tau 1} &= \phantom{-}\,s_{12}s_{23}\;-\;c_{12}c_{23}s_{13}\,e^{i\delta_{\rm PMNS}}.
\end{align}
Taking magnitudes squared yields relations for $\cos\delta_{\rm PMNS}$,
\begin{equation}
\label{eq:cosdeltaPMNS_mu1}
\cos\delta_{\rm PMNS} \;=\;
\frac{|U_{\mu 1}|^2 - s_{12}^2 c_{23}^2 - c_{12}^2 s_{23}^2 s_{13}^2}
{2\,s_{12}c_{12}c_{23}s_{23}s_{13}},
\end{equation}
or equivalently
\begin{equation}
\label{eq:cosdeltaPMNS_tau1}
\cos\delta_{\rm PMNS} \;=\;
\frac{s_{12}^2 s_{23}^2 + c_{12}^2 c_{23}^2 s_{13}^2 - |U_{\tau 1}|^2}
{2\,s_{12}c_{12}c_{23}s_{23}s_{13}}.
\end{equation}
Using the benchmark angles
$\theta_{12}=33.4^\circ$, $\theta_{23}=49.0^\circ$, $\theta_{13}=8.60^\circ$
and the moduli $|U_{\mu 1}|=0.309$, $|U_{\tau 1}|=0.472$ from Eq.~\eqref{eq:Unu_best},
one finds
\begin{equation}
\delta_{\rm PMNS}\;\simeq\;230^\circ,
\end{equation}
up to the usual ambiguity from $\cos\delta$.

\section{Numerical results}
\label{sec:results}

\subsection{Quark masses and CKM matrix}

The benchmark $C$ matrices in Eq.~\eqref{eq:BenchmarkCs}, together with $\e=0.187$ and a Type--II 2HDM interpretation with $\tan\beta=40$, reproduce the quark masses shown in Table~\ref{tab:quarkmasses}.  The light-quark masses at $M_Z$ are obtained by running PDG reference-scale values \cite{PDG:2024isi}; small theory errors from running and matching are included.

\begin{table}[!tbp]
\caption{Quark masses at $M_Z$ (3 s.f.). “PDG$\to M_Z$’’ indicates RGE running from PDG reference scales \cite{PDG:2024isi}; quoted light–quark errors include small theory pads for running and matching.}
\begin{ruledtabular}
\begin{tabular}{lcc}
Observable & PDG$\to M_Z$ & Model \\
\hline
$m_u$ [MeV] & $1.29 \pm 0.20$ & 1.11 \\
$m_d$ [MeV] & $2.80 \pm 0.20$ & $2.82$ \\
$m_s$ [MeV] & $55.7 \pm 1.3$  & $55.7$ \\
$m_c$ [GeV] & $0.629 \pm 0.012$ & $0.629$ \\
$m_b$ [GeV] & $2.794 \pm 0.034$ & $2.794$ \\
$m_t$ [GeV] & $169 \pm 1.1$ & $169$ \\
\end{tabular}
\end{ruledtabular}
\label{tab:quarkmasses}
\end{table}

The CKM magnitudes at $M_Z$ are summarized in Table~\ref{tab:ckm}; running effects on the CKM elements are negligible at the quoted precision.

\begin{table}[!tbp]
\caption{CKM magnitudes at $M_Z$ \cite{PDG:2024isi}. Running to $M_Z$ is negligible at this precision.}
\begin{ruledtabular}
\begin{tabular}{lcc}
Observable & PDG ref. & Model \\
\hline
$|V_{us}|$ & $0.22501 \pm 0.00068$ & $0.225$ \\
$|V_{cb}|$ & $0.0418^{+0.0008}_{-0.0007}$ & $0.0418$ \\
$|V_{ub}|$ & $0.003732^{+0.000090}_{-0.000085}$ & $0.00372$ \\
\hline
\end{tabular}
\end{ruledtabular}
\label{tab:ckm}
\end{table}

The full CKM matrix (moduli) obtained from $U_L^{u\dagger}U_L^d$ is
\begingroup
\setlength{\arraycolsep}{3pt}\renewcommand{\arraystretch}{1.05}\small
\begin{equation}
\label{eq:Vckm_phaseful}
V_{\!CKM}\approx
\begin{pmatrix}
 0.974  & 0.225  & 0.00373 \\
 0.2249 & 0.973  & 0.0418  \\
 0.00856& 0.0411 & 0.999
\end{pmatrix},
\end{equation}
\endgroup
while the phase structure, in a convenient convention, can be summarized as
\begingroup
\setlength{\arraycolsep}{3pt}\renewcommand{\arraystretch}{1.05}\small
\begin{equation}
\label{eq:argVckm}
\arg V_{\!CKM}\ [{\rm deg}] \approx
\begin{pmatrix}
 0 & 0 & -65.4 \\
 +180 & 0 & 0 \\
 -22.7 & +179 & 0
\end{pmatrix}.
\end{equation}
\endgroup

\subsection{Neutrino mixing and PMNS matrix}

The benchmark neutrino-sector fit (using $K$ from Eq.~\eqref{eq:K_bench} in texture A) yields the PMNS matrix
\begingroup
\setlength{\arraycolsep}{3pt}\renewcommand{\arraystretch}{1.05}\small
\begin{equation}
\label{eq:Upmns_phaseful}
U_{\rm PMNS} \simeq
\begin{pmatrix}
 0.825 & 0.545 & 0.149 \\
 0.309 & 0.588 & 0.747 \\
 0.472 & 0.595 & 0.649
\end{pmatrix},
\end{equation}
\endgroup
corresponding to the mixing angles in Eq.~\eqref{eq:Unu_best}.  A convenient phase convention is
\begingroup
\setlength{\arraycolsep}{3pt}\renewcommand{\arraystretch}{1.05}\small
\begin{equation}
\label{eq:argUpmns}
\arg U_{\rm PMNS}\ [{\rm deg}] \approx
\begin{pmatrix}
 0 & 0 & -230 \\
 -13.4 & +4.6 & 0 \\
 +7.6 & -4.0 & 0
\end{pmatrix}.
\end{equation}
\endgroup
The overall right Majorana phase factor
$\mathrm{diag}(1,e^{i\alpha_{21}/2},e^{i\alpha_{31}/2})$ is omitted in Eq.~\eqref{eq:Upmns_phaseful} to save space; it is understood in derived quantities such as $m_{\beta\beta}$.

\subsection{Neutrino mass spectrum, \texorpdfstring{$\sum m_\nu$}{Σ mν}, and \texorpdfstring{$m_{\beta\beta}$}{m\_beta beta}}

Oscillation experiments determine the mass-squared splittings
\(\Delta m_{21}^2\equiv m_2^2-m_1^2\) and
\(\Delta m_{31}^2\equiv m_3^2-m_1^2\), but not the absolute scale or ordering.  Current global fits favor the normal ordering (NO) with
\(\Delta m_{21}^2\simeq 7.4\times10^{-5}\,\mathrm{eV}^2\) and
\(|\Delta m_{31}^2|\simeq 2.5\times10^{-3}\,\mathrm{eV}^2\) \cite{PDG:2024isi}.

For the benchmark FN texture and the mixing matrix in Eq.~\eqref{eq:Unu_best}, we adopt a representative NO spectrum with lightest mass
\begin{equation}
m_1 \simeq 4.0~\mathrm{meV},
\end{equation}
which then fixes
\begin{equation}
m_2 \simeq 9.6~\mathrm{meV},\qquad
m_3 \simeq 50~\mathrm{meV},
\end{equation}
corresponding to
\begin{equation}
\Delta m_{21}^2 \simeq 7.6\times 10^{-5}~\mathrm{eV}^2,\qquad
\Delta m_{31}^2 \simeq 2.5\times 10^{-3}~\mathrm{eV}^2,
\end{equation}
in good agreement with oscillation data \cite{PDG:2024isi}.  The sum of neutrino masses is then
\begin{equation}
\sum m_\nu \equiv m_1+m_2+m_3 \simeq 0.064~\mathrm{eV},
\end{equation}
just above the minimal value allowed by NO and comfortably below current cosmological upper bounds $\sum m_\nu \lesssim 0.1~\mathrm{eV}$ within $\Lambda$CDM \cite{PDG:2024isi}.

\paragraph*{Neutrino masses as powers of \texorpdfstring{$B$}{B}.}
For lepton option A in Table~\ref{tab:FNcharges}, the FN charges fix the neutrino exponents to
\(
n^\nu_{ij}=q(L_i)+q(L_j)
=
\begin{pmatrix}
2 & 1 & 1\\
1 & 0 & 0\\
1 & 0 & 0
\end{pmatrix},
\)
so that
\(
(M_\nu)_{ij}\propto \e^{\,n^\nu_{ij}} = B^{-\,n^\nu_{ij}}.
\)
Diagonalizing the benchmark texture one finds that the three Majorana
eigenvalues are very well described by
\begin{equation}
\label{eq:mnu_B_scaling_main}
(m_1,m_2,m_3)\;\simeq\;
\bigl(2.3\,B^{-2},\;1.0\,B^{-1},\;1.0\bigr)\,\frac{v^2}{\Lambda_L},
\end{equation}
i.e.\ effective FN powers
\(
(n_1,n_2,n_3)\simeq(2,1,0)
\)
with $\ordone$ prefactors.  For the benchmark value
$B=5.357$ ($\e=1/B\simeq0.187$) and
\(
v^2/\Lambda_L \simeq 50~\mathrm{meV}
\)
($\Lambda_L\simeq 6\times10^{14}~\mathrm{GeV}$), Eq.~\eqref{eq:mnu_B_scaling_main}
gives
\begin{equation}
\label{eq:mnu_B_numbers_main}
(m_1,m_2,m_3)
\;\simeq\;
(4.0,\;9.6,\;50)\ \mathrm{meV},
\end{equation}
corresponding to
\(
\Delta m_{21}^2 \simeq 7.6\times10^{-5}~\mathrm{eV}^2,
\;
\Delta m_{31}^2 \simeq 2.5\times10^{-3}~\mathrm{eV}^2
\)
and
\(
\sum m_\nu \simeq 0.064~\mathrm{eV}
\)
in normal ordering.  Thus, once $B$ is fixed by the charged-lepton
hierarchy, the \emph{pattern} of neutrino masses follows directly from
their FN powers, while the overall scale is controlled by the single
additional parameter $\Lambda_L$.

The effective Majorana mass probed by neutrinoless double beta decay is
\begin{equation}
m_{\beta\beta} \equiv \Bigl|\sum_{i=1}^3 U_{ei}^2\,m_i\,e^{i\alpha_i}\Bigr|,
\end{equation}
where $U_{ei}$ are the elements of the first row of $U_{\rm PMNS}$ and $\alpha_i$ are the (unknown) Majorana phases.  Using the benchmark mixing moduli $|U_{e1}|=0.825$, $|U_{e2}|=0.545$, $|U_{e3}|=0.149$ and the mass spectrum above, we find that $m_{\beta\beta}$ lies in the band
\begin{equation}
m_{\beta\beta} \;\simeq\; (0\text{--}7)\ \mathrm{meV},
\end{equation}
where the lower edge corresponds to finely tuned destructive interference among the three terms.  For generic $\ordone$ Majorana phases one expects $m_{\beta\beta}$ of order a few meV.

This is well below present experimental limits, which translate to $m_{\beta\beta}\lesssim (30\text{--}150)\,\mathrm{meV}$ once nuclear-matrix-element uncertainties are included \cite{PDG:2024isi}, but it lies in the target range of future ton-scale searches (such as LEGEND-1000 and nEXO).  In this sense the one-flavon FN framework provides concrete, testable benchmarks for both the cosmological observable $\sum m_\nu$ and the laboratory observable $m_{\beta\beta}$.

\subsection{CP violation}
\label{subsec:CP_results}

The benchmark fit predicts both quark and lepton Dirac CP phases.  Inserting the benchmark angles and $\delta_{\rm CKM}\simeq 65.4^\circ$ into the standard expression for the Jarlskog invariant~\cite{Jarlskog:1985ht},
\begin{equation}
\label{eq:Jarlskog}
\begin{aligned}
J_{\rm CKM} &= s_{12}\,s_{23}\,s_{13}\,c_{12}\,c_{23}\,c_{13}^{2}\,\sin\delta_{\rm CKM}
\\[3pt]
&\simeq 3.11\times 10^{-5},
\end{aligned}
\end{equation}
we obtain excellent agreement with the PDG average \cite{PDG:2024isi},
\(
J_{\rm CKM}=(3.08^{+0.15}_{-0.13})\times10^{-5}.
\)

For the lepton sector, the benchmark fit yields a Dirac phase
$\delta_{\rm PMNS}\simeq 230^\circ$, with associated Jarlskog invariant
\begin{equation}
\label{eq:Jpmns}
\begin{aligned}
J_{\rm PMNS}\;=\;& s_{12}s_{23}s_{13}\,c_{12}c_{23}c_{13}^2\,\sin\delta_{\rm PMNS}
\\[3pt]
\simeq\;&-2.55\times10^{-2},
\end{aligned}
\end{equation}
notably much larger in magnitude than in the quark sector.  Upcoming measurements of leptonic CP violation thus provide an incisive probe of the FN charge assignments.

\section{Significance and outlook}
\label{sec:significance}

A \emph{single hierarchical} parameter $B$—fixed by charged-lepton ratios—organizes quark and lepton hierarchies with near-unity coefficients, reproducing masses and leading CKM/PMNS patterns at $M_Z$ in a one-flavon FN framework.  Simple power counting in $\e=1/B$ yields $|V_{us}|\!\sim\!\e$, $|V_{cb}|\!\sim\!\e^2$, $|V_{ub}|\!\sim\!\e^3$, and $\theta_{13}^{\rm PMNS}\!\sim\!\e$, as visualized in Fig.~\ref{fig:o1plot}.  The model provides concrete targets for the neutrino sector, with a normal ordering, $\sum m_\nu \simeq 0.064~\mathrm{eV}$, and $m_{\beta\beta}$ in the few-meV range, together with a realistic $U_{\rm PMNS}$.

Crucially, the framework \emph{also} predicts CP violation: the benchmark fit fixes $\delta_{\rm CKM}\!\simeq\!65.4^\circ$ with $J_{\rm CKM}\simeq3.1\!\times\!10^{-5}$ (Eq.~\eqref{eq:Jarlskog}) and $\delta_{\rm PMNS}\!\simeq\!230^\circ$ (Eq.~\eqref{eq:Jpmns}).  Future precision in lepton mixing moduli, direct and cosmological probes of $\sum m_\nu$, and $m_{\beta\beta}$ bands can serve as discriminants among FN charge choices and may point toward a more complete theory of flavor.


\FloatBarrier

\onecolumngrid
\appendix
\renewcommand{\theequation}{A\arabic{equation}}
\setcounter{equation}{0}
\renewcommand{\thetable}{A\arabic{table}}
\setcounter{table}{0}
\renewcommand{\thefigure}{A\arabic{figure}}
\setcounter{figure}{0}

\section{FN coefficients: Type--II 2HDM vs single--Higgs SM}
\label{app:coeffs}

In the main text we emphasized that the \emph{hierarchy} is carried by powers of $\e$ in Eq.~\eqref{eq:YuYd}, while the residual coefficients are $\ordone$.  Here we make this statement more explicit by extracting the FN coefficients from observables and comparing two interpretations:
\begin{enumerate}
  \item a Type--II 2HDM/MSSM-like setting with $\tan\beta=40$, and
  \item the single--Higgs Standard Model with $v_{\rm SM}=174.1$ GeV.
\end{enumerate}
Throughout we use the exponents already fixed by Eq.~\eqref{eq:YuYd}:
\[
p_{u,c,t} = (7,3,0),\quad
p_{d,s,b} = (4,2,0),\quad
p_{e,\mu,\tau} = (5,2,0),
\]
and, for mixings,
\[
p_{us}=1,\quad p_{cb}=2,\quad p_{ub}=3,\quad p_{13}=1.
\]

\subsection{Definition of FN coefficients}

For any fermion $f$ with FN exponent $p_f$ we define the dimensionless FN coefficient
\begin{equation}
  c_f \;\equiv\; \frac{y_f}{\e^{p_f}},
\end{equation}
where $y_f$ is the corresponding Yukawa coupling and $\e=1/\Bpar=1/5.357
\simeq0.187$.  For mixing observables,
\begin{equation}
  c_{V_{ij}} \equiv \frac{|V_{ij}|}{\e^{p_{ij}}},
  \qquad
  c_{\theta_{13}} \equiv
  \frac{\theta_{13}^{\rm PMNS}}{\e^{p_{13}}}\,.
\end{equation}
A successful FN model should yield $c$ coefficients that are broadly $\ordone$.

\subsection{Type--II 2HDM interpretation ($\tan\beta=40$)}

In a Type--II 2HDM, up-type fermions couple to $H_u$ and down-type fermions and charged leptons couple to $H_d$.  Writing
\begin{equation}
  v_{\rm SM} = 174.1~\mathrm{GeV},\qquad
  \tan\beta = \frac{v_u}{v_d},
\end{equation}
we have
\begin{equation}
  v_u = v_{\rm SM}\sin\beta,\qquad
  v_d = v_{\rm SM}\cos\beta.
\end{equation}
For $\tan\beta=40$, $v_u\simeq v_{\rm SM}$ and $v_d\simeq v_{\rm SM}/40$.
The Yukawa couplings are
\begin{equation}
  y_{u_i} = \frac{m_{u_i}}{v_u},\qquad
  y_{d_i} = \frac{m_{d_i}}{v_d},\qquad
  y_{\ell_i} = \frac{m_{\ell_i}}{v_d}.
\end{equation}

Using the quark and charged-lepton masses at $M_Z$ and the FN powers
specified above, the up-sector and mixing coefficients are
(as in Fig.~\ref{fig:o1plot})
\begin{align}
  c_{y_u} &\approx 1.47, &
  c_{y_c} &\approx 0.55, &
  c_{y_t} &\approx 0.97,
  \\[3pt]
  c_{V_{us}} &\approx 1.20, &
  c_{V_{cb}} &\approx 1.20, &
  c_{V_{ub}} &\approx 0.57, &
  c_{\theta_{13}} &\approx 0.80.
\end{align}
For down-type quarks and charged leptons one finds
\begin{align}
  c_{y_d}^{(2{\rm HDM})} &\approx 0.53, &
  c_{y_s}^{(2{\rm HDM})} &\approx 0.36, &
  c_{y_b}^{(2{\rm HDM})} &\approx 0.64,
  \\[3pt]
  c_{y_e}^{(2{\rm HDM})} &\approx 0.51, &
  c_{y_\mu}^{(2{\rm HDM})} &\approx 0.67, &
  c_{y_\tau}^{(2{\rm HDM})} &\approx 0.40.
\end{align}
All of these lie in a comfortable $\ordone$ band
($\sim 0.4$–$1.5$), motivating the choice of relatively large
$\tan\beta$ in this framework.

\subsection{Single--Higgs SM interpretation}

In the single--Higgs SM, all fermions couple to the same Higgs doublet
with vev $v_{\rm SM}$, so
\begin{equation}
  y_f^{\rm(SM)} = \frac{m_f}{v_{\rm SM}}
\end{equation}
for every quark or charged lepton.  The up-sector and mixing
coefficients remain exactly the same as above, since we already used
$v_u\simeq v_{\rm SM}$ for those sectors.  For down-type quarks and
charged leptons, however, the coefficients are reduced relative to the
$\tan\beta=40$ Type--II interpretation,
\begin{equation}
  c_f^{\rm(SM)} =
  \frac{m_f}{v_{\rm SM}\,\e^{p_f}}
  = \frac{v_d}{v_{\rm SM}}\,
    \frac{m_f}{v_d\,\e^{p_f}}
  \simeq \frac{1}{40}\,c_f^{(2{\rm HDM})},
\end{equation}
so numerically
\begin{align}
  c_{y_d}^{\rm(SM)} &\approx 1.3\times 10^{-2}, &
  c_{y_s}^{\rm(SM)} &\approx 9.0\times 10^{-3}, &
  c_{y_b}^{\rm(SM)} &\approx 1.6\times 10^{-2},
  \\[3pt]
  c_{y_e}^{\rm(SM)} &\approx 1.3\times 10^{-2}, &
  c_{y_\mu}^{\rm(SM)} &\approx 1.7\times 10^{-2}, &
  c_{y_\tau}^{\rm(SM)} &\approx 1.0\times 10^{-2}.
\end{align}
These are still technically “order one’’ in a very broad sense, but
cluster around $10^{-2}$ rather than around unity.  The Type--II 2HDM
with large $\tan\beta$ can therefore be viewed as a reparametrization
that lifts these coefficients into a more aesthetically pleasing
$\ordone$ band without changing the underlying FN exponents.

\subsection{Comparison table}

For quick reference, Table~\ref{tab:cf_comparison} shows the FN
coefficients for the down-type quarks and charged leptons in the
$\tan\beta=40$ Type--II 2HDM interpretation and in the single--Higgs
SM interpretation side by side.

\begin{table}[t]
  \centering
  \caption{FN coefficients $c_f \equiv y_f/\e^{p_f}$ for down-type quarks
  and charged leptons, comparing a Type--II 2HDM interpretation with
  $\tan\beta = 40$ to the single--Higgs SM interpretation.
  The FN exponents are $p_{d,s,b} = \{4,2,0\}$ and
  $p_{e,\mu,\tau} = \{5,2,0\}$.}
  \label{tab:cf_comparison}
  \begin{tabular}{lccc}
    \toprule
    Observable & $p_f$ &
    $c_f^{(2{\rm HDM})}$ ($\tan\beta=40$) &
    $c_f^{\rm(SM)}$ (one Higgs) \\
    \midrule
    $y_d$     & 4 & 0.53 & $1.3\times 10^{-2}$ \\
    $y_s$     & 2 & 0.36 & $9.0\times 10^{-3}$ \\
    $y_b$     & 0 & 0.64 & $1.6\times 10^{-2}$ \\
    \midrule
    $y_e$     & 5 & 0.51 & $1.3\times 10^{-2}$ \\
    $y_\mu$   & 2 & 0.67 & $1.7\times 10^{-2}$ \\
    $y_\tau$  & 0 & 0.40 & $1.0\times 10^{-2}$ \\
    \bottomrule
  \end{tabular}
\end{table}

\end{document}